\documentclass[12pt]{article}
\usepackage{amsmath}
\usepackage{subcaption}
\usepackage{enumerate}
\usepackage{natbib}
\usepackage{xcolor}
\usepackage{appendix}
\usepackage{url}
\usepackage{multirow}
\usepackage{amsthm} 

\usepackage{tablefootnote}
\usepackage{longtable}
\usepackage{threeparttable}
\usepackage{bm}
\usepackage[ruled,vlined]{algorithm2e}
\usepackage{caption}
\usepackage{placeins}
\usepackage{lscape}

\usepackage{graphicx,booktabs}

\newtheorem{theorem}{Theorem}

\newtheorem{lemma}[theorem]{Proposition}

\newcommand{\blind}{1}

\begin{document}

\def\spacingset#1{\renewcommand{\baselinestretch}%
{#1}\small\normalsize} \spacingset{1}

\if1\blind
{
  \title{\bf Estimating and Correcting Degree Ratio Bias in the Network Scale-up Method}
  \author{Ian Laga, Jessica P. Kunke, Tyler McCormick, Xiaoyue Niu}
  \maketitle
} \fi

\if0\blind
{
  \bigskip
  \bigskip
  \bigskip
  \begin{center}
    {\LARGE\bf Estimating and Correcting Degree Ratio Bias in the Network Scale-up Method}
\end{center}
  \medskip
} \fi

\bigskip
\begin{abstract}
    The Network Scale-up Method (NSUM) uses social networks and answers to ``How many X's do you know?" questions to estimate sizes of groups excluded by standard surveys. This paper addresses the bias caused by varying average social network sizes across populations, commonly referred to as the degree ratio bias. This bias is especially important for marginalized populations like sex workers and drug users, where members tend to have smaller social networks than the average person. We show how the degree ratio affects size estimates and provide a method to estimate degree ratios without collecting additional data. We demonstrate that our adjustment procedure improves the accuracy of NSUM size estimates using simulations and data from two data sources.
\end{abstract}

\noindent%
\vfill

\newpage

\spacingset{2} 

\section{Introduction}
\label{sec:intro}

The Network Scale-up Method (NSUM) has emerged as a popular and efficient way to estimate the size of hard-to-reach populations such as female sex workers, drug users, and men who have sex with men. These hard-to-reach populations are of critical importance to solving several global health problems, including meeting UNAIDS HIV-related targets \citep{unaids2021inequality}. These populations are at a higher risk of contracting and spreading HIV than the general population while simultaneously suffering from marginalization and negative social stigma.

The NSUM estimates the size of these populations using survey questions of the form ``How many X's do you know," where X includes both subpopulations with known sizes and subpopulations of interest with unknown sizes, such as female sex workers \citep{bernard1989estimating}. These survey responses are known as aggregated relational data (ARD). While some research on ARD concerns the estimation of network structures \citep{breza2020using}, we focus on the role ARD play in estimating hard-to-reach subpopulation sizes.

Previous researchers have proposed several modeling improvements to better capture the complexity of the underlying aggregated relational data, including those by \cite{zheng2006many}, \cite{maltiel2015estimating}, \cite{teo2019estimating}, and \cite{laga2021correlated}. These approaches aim to either better understand underlying network properties or improve population size estimates from NSUM models by incorporating underlying network properties into the model.

This work focuses on the NSUM subpopulation size estimator proposed in \cite{killworth1998estimation}, which we refer to as the basic scale-up estimator (see \cite{mccormick2020network} or \cite{laga2021thirty} for a comprehensive review). The basic scale-up estimator is currently the most commonly used NSUM estimator. \cite{killworth1998estimation} assume the ARD come from the following distribution:
\begin{equation}
\label{eq:binomial}
    y_{ik} \sim \text{Binomial}\left(d_i, N_k/N\right),
\end{equation}
where $y_{ik}$ denotes the number of people respondent $i$ reports knowing in subpopulation $k$, $d_i$ is the degree, or social network size, of respondent $i$, $N$ is the size of the total population, and $N_k$ is the size of subpopulation $k$. In a survey with $K$ subpopulations, we let ``known subpopulations'' refer to the $L$ subpopulations where $N_k$ is known, and ``unknown subpopulations'' refer to the $K - L$ subpopulations where $N_k$ is unknown. For simplicity, we assume there is only one unknown subpopulation, $H$, although in general there may be more than one. Assuming that we know the true degrees, $d_i$, or are able to consistently estimate them from the ARD, the subpopulation size estimate for $H$ is given by
\begin{equation*}
    \hat{N}_H = N\frac{\sum_{i=1}^n y_{iH}}{\sum_{i=1}^n d_i},
\end{equation*}
which is the maximum likelihood estimator for $N_H$ from Equation \eqref{eq:binomial} conditional on the $d_i$.

The basic scale-up estimator is subject to a variety of biases, including when respondents are more or less likely to know people from certain populations (barrier effects), do not know everything about their social contacts (transmission error), or cannot accurately recall everyone in their social network (recall error). We focus on the degree ratio error introduced by different subpopulations having different average network sizes. Specifically, the degree ratio for subpopulation $k$ is the ratio between the average degree of members of subpopulation $k$ to the average degree of individuals who may be included as respondents in the ARD survey. \cite{feehan2016generalizing} propose a generalized scale-up estimator and show that their estimator is equal to the basic scale-up estimator multiplied by three adjustment factors, one of which is the degree ratio. While the authors propose several approaches to correct for these factors, correcting for the degree ratio typically requires collecting additional survey data directly from the subpopulation. For example, \cite{salganik2011game} created the game of contacts, which involves interviewing members of the hard-to-reach population. While their original motivation for the game of contacts was the estimate the transmission error, it may also be used to estimate the degree ratio. Alternatively, \cite{feehan2016generalizing} propose collecting additional ARD from the hard-to-reach population to estimate the degree ratio.

Failing to account for the degree ratio can significantly bias NSUM subpopulation size estimates. \cite{shelley1995knows} found that HIV positive respondents and respondents who were dialysis patients had networks which were only about 2/3 the size of those of the average respondent in their survey. Therefore, given perfect responses to ARD questions, the basic scale-up estimator would estimate the size of these two subpopulations to be about 2/3 of the true size. The degree ratio may be more influential for even more stigmatized populations like sex workers or for more social populations like priests and doctors.

In this work, we propose a simple approach to estimate and correct for the degree ratio based on the linear relationship between respondents' social network sizes and the number of people they know in different subpopulations. Our approach conveniently relies on only the original ARD, allowing researchers to obtain more accurate size estimates without collecting additional data like those needed for the game of contacts and the generalized scale-up estimator.

The rest of this paper is organized as follows. First, Section \ref{sec:improve_background} provides additional background information about the degree ratio and presents the bias of the basic scale-up estimator under certain conditions. Then, in Section \ref{sec:degree_ratio}, we introduce our approach to estimate the degree ratio using only the original ARD responses. We apply this approach to both simulated (Section \ref{sec:simulation}) and real (Section \ref{sec:realdata}) ARD surveys. Finally, we close with a discussion in Section \ref{sec:improve_discussion}.

\section{Background}
\label{sec:improve_background}

We first review model properties of the basic scale-up estimator. \cite{feehan2016generalizing} show that the basic scale-up estimator is equivalent to their generalized scale-up estimator only when multiplied by three adjustment factors, one of which involves the degree ratio. The degree ratio adjustment factor arises because some populations have larger or smaller social network sizes on average than other populations. The authors define the degree ratio, $\delta_F$, as
\begin{equation*}
    \delta_F = \frac{\text{avg \# connections from a member of H to F}}{\text{avg \# connections from a member of F to the rest of F}} = \frac{\bar{d}_{H,F}}{\bar{d}_{F,F}},
\end{equation*}
where $F$ refers to the frame population, the collection of individuals who may be included as respondents in the ARD survey, and $H$ refers to the hidden or unknown subpopulation. Thus, if the degree ratio is 0.5 (i.e. there are only half as many links per member of $H$ from $H$ to $F$ as there are per member of $F$ from $F$ to $F$), then the basic scale-up estimator is one half the size of the generalized scale-up estimator. Since the basic scale-up estimator implicitly assumes the average degrees of all subpopulations, both known and unknown, are identical, the estimator misattributes the small number of links to a small subpopulation size, rather than to small degrees. In order to estimate these adjustment factors, \cite{feehan2016generalizing} propose collecting an additional ARD survey given to members of the hard-to-reach population, to collect what they call enriched ARD.

While we recognize the utility of enriched ARD, there are three significant limitations. First, enriched ARD is often prohibitively expensive to collect. The low cost and easy implementation of the NSUM are two of its key benefits. Collecting enriched ARD therefore undermines this advantage since only well-funded studies will be able to collect the additional data. Second, it is impossible to collect enriched ARD on impossible-to-reach subpopulations such as individuals who died in an earthquake. Finally, it is inconvenient or impossible to collect enriched ARD for previous ARD studies, so the methods proposed in \cite{feehan2016generalizing} can only naturally be used for ARD moving forward. To correct for the biases in existing ARD surveys that did not already collect enriched ARD, users must either assume an adjustment factor for the degree ratio and construct confidence intervals using the rescaled bootstrap procedure (as proposed by \cite{feehan2016generalizing}), or find and survey a similar contemporary population and assume the behavior of the two subpopulations are similar. Instead, we propose the first method to estimate the degree ratio using only the original ARD, allowing researchers to easily correct for bias introduced by the degree ratio.

We present two related findings connecting the bias of the basic scale-up estimator to the degree ratio. For the following results, we assume perfect link reporting (i.e. no transmission error or recall error), that the respondents represent a simple random sample $S$ of size $n$ from the entire population of size $N$, and that the frame population $F$ is the entire population, where $H$ is included in $F$. In this case, the inclusion probability for each respondent $i$ is $\pi_i = n / N$. We consider two estimators, where either (1) the $d_i$ are fixed and known, or (2) the $d_i$ are estimated using the $L$ known subpopulations. In the first case, we can represent the basic scale-up estimator as
\begin{equation}
    \hat{N}_{H,1} = \frac{\sum_{i \in S} (y_{iH}/\pi_i)}{\frac{1}{N} \sum_{i\in S} (d_i/\pi_i)},
\label{eq:mle_known}
\end{equation}
while the second case includes the estimation of $d_i$, given by
\begin{equation}
    \hat{N}_{H,2} = \frac{\sum_{i \in S} (y_{iH}/\pi_i)}{\frac{1}{N} \sum_{i\in S} \left[\left(\sum_{k=1}^L y_{ik} / \sum_{k=1}^L N_k \right)/\pi_i \right]}.
\label{eq:mle_unknown}
\end{equation}
Using these estimators, we present the following propositions, where Proposition 1 is a special-case result from \cite{feehan2016generalizing} adapted here for completeness, and the proof for Proposition 2 is shown in Appendix \ref{sec:proofs}.

\begin{lemma}
{\normalfont  \textbf{Adapted from \cite{feehan2016generalizing}}.} Consider the size estimate $\hat{N}_{H,1}$ in Equation \eqref{eq:mle_known}, obtained from a survey with perfect link reporting and from a simple random sample of respondents. Then given known degrees $d_i$, the bias of the unknown size estimate is approximately given by
\begin{equation*}
    \text{Bias}(\hat{N}_{H,1}) \approx N_H \left(\frac{\bar{d}_H}{\bar{d}_F} - 1 \right),
\end{equation*}
where $\bar{d}_H$ denotes the average degree of individuals in the hidden subpopulation and $\bar{d}_F$ denotes the average degree of individuals in the frame population.
\end{lemma}

\begin{lemma}
Consider the size estimate $\hat{N}_{H,2}$ in Equation \eqref{eq:mle_unknown}, obtained from a survey with perfect link reporting and from a simple random sample of respondents. Then given that the $d_i$ are estimated using the $L$ known subpopulations in the survey, the bias of the unknown size estimate is approximately given by
\begin{equation*}
    \text{Bias}(\hat{N}_{H,2}) \approx N_H \left(\frac{\bar{d}_H \sum_{k=1}^L N_k}{ \sum_{k=1}^L \bar{d}_k N_k } - 1 \right),
\end{equation*}
where $\bar{d}_H$ denotes the average degree of individuals in the hidden subpopulation and $\bar{d}_k$ denotes the average degree of individuals in subpopulation $k$.
\end{lemma}

These results show that when the true degrees are known, the bias depends only on the true subpopulation size of $H$ and the ratio of the average degrees between the unknown subpopulation and the frame population, while the bias of the estimator when the degrees are also estimated additionally depends on the remaining known subpopulation sizes and the average degrees of individuals in each known subpopulation size. Proposition 2 also shows that the accuracy of the unknown size estimate depends on the specific relationship between the average degrees in subpopulations and the sizes of those subpopulations, and relatively large or small subpopulations will introduce more bias when paired with relatively large or small average degrees, respectively.

\section{Degree Ratio Adjustment}
\label{sec:degree_ratio}

Here we propose a method to correct for the bias introduced by the degree ratio in the basic scale-up estimator. It would be sufficient to know $\bar{d}_k$ for all known and unknown subpopulations. However, these average degrees are unknown, making a direct approach impossible. Furthermore, a primary advantage of the NSUM is avoiding contacting members of hard-to-reach populations. Thus, our goal is to estimate $\delta_k$, the degree ratio for subpopulation $k$, using existing ARD data to produce an adjustment factor for $\hat{N}_k$.

For the remainder of this paper, we let $\delta_k = \bar{d}_k / \bar{d}_F$ represent the degree ratio for subpopulation $k$, where we depart from the original notation from \cite{feehan2016generalizing} to emphasize that the degree ratio (i) exists for both the subpopulations with known size and the hard-to-reach subpopulations, and (ii) varies across subpopulation. 

The basic assumption of our approach is that the proportion of an individual's social network that belongs to group $k$ depends on the individual's degree. To incorporate this assumption, we modify Equation \eqref{eq:binomial} such that
\begin{equation}
    y_{ik} \sim \text{Binomial}\left(d_i, \frac{N_k}{N} f_k(d_i)\right),
    \label{eq:binomial_bias}
\end{equation}
where $f_k(d_i)$ is of the form $f_k(d_i) = a + g(d_i) c_k$ for each $k$ across all values of $d_i$, $a \neq 0$, and $g(d_i)$ is any finite-valued function of $d_i$ that does not depend on $k$. Additionally, $c_k$ is a group-specific term that controls how an individual's degree affects the probability of knowing people from group $k$. Necessarily, $f_k(d_i)$ is limited to functions such that $N_k f_k(d_i) / N$ is between 0 and 1. The form of $g(d_i)$ does not need to be known, but must be the same for all $k$.

In the context of NSUM, an example of a reasonable $f_k(d_i)$ is a mean-one function of the form
\begin{equation}
    f_k(d_i) = 1 + \left(d^p_i - \frac{1}{n}\sum_{i=1}^n d^p_i\right) c_k = 1 + \left(d^p_i - \bar{d^p}\right) c_k,
    \label{eq:fk}
\end{equation}
where $d_i^p$ represents $d_i$ to the $p^{th}$ power. The form of $f_k(d_i)$ is general enough to account for many realistic situations. First, $c_k$ may be positive or negative, leading to respondents with larger degrees having higher binomial probabilities when $c_k > 0$, and vice versa when $c_k < 0$. Second, the power $p$ controls how quickly departures from $\bar{d^p}$ affect the probabilities. The difference between biases for varying values of $p$ depends on the specific degree distribution of the respondents. Furthermore, the choice of $p$ also controls what degree corresponds to $f_k(d_i) = 1$.

Given the above, we have the following result.
\begin{lemma}
\label{lemma:binomial}
Consider aggregated relational data generated from the likelihood defined by Equation \eqref{eq:binomial_bias} for any $f_k(d_i)$ of the form $f_k(d_i) = a + g(d_i) c_k$, where $a \neq 0$, $g(d_i)$ is any finite-valued function of $d_i$ that does not depend on $k$, and $f_k(d_i)$ leads to a valid likelihood for all $k$. Then given known degrees $d_i$ and the estimator in Equation \eqref{eq:mle_known}, the bias of $\hat{N}_k$ is given by
\begin{equation*}
    \text{Bias}(\hat{N}_k) \approx N_k \left(\frac{\sum_{i=1}^n d_i f_k(d_i)}{\sum_{i=1}^n d_i} - 1 \right).
\end{equation*}
Furthermore, there exists some $\gamma_0$ and $\gamma_1$, which are independent of $k$, such that
\begin{equation*}
     E\left(\frac{N_k}{\hat{N}_k}\right) \approx E\left(\gamma_0 + \gamma_1 \left( \frac{\sum_{i=1}^n (d_i - \bar{d}_i)y_{ik}}{\frac{1}{n}\left(\sum_{i=1}^n y_{ik}\right) \left(\sum_{i=1}^n (d_i - \bar{d}_i)^2\right)} \right) \right).
\end{equation*}
\end{lemma}

The proof of Proposition \ref{lemma:binomial} is in Appendix \ref{sec:proof3}. Proposition \ref{lemma:binomial} provides a specific form of the degree ratio under our assumed binomial likelihood. From Proposition 1, we have $E(\hat{N}_k / N_k) \approx \delta_k$ and we show in Appendix \ref{sec:proof3} that $E(N_k / \hat{N}_k) \approx 1 / \delta_k$. Based on Proposition \ref{lemma:binomial}, we need only to first estimate $\gamma_0$ and $\gamma_1$ to then estimate $1/\delta_k$. After estimating $1/\delta_k$, we can create an approximately unbiased estimator, $N_k^{adj} = \hat{N}_k / \hat{\delta}_k$. We note that while the results hold only for a fixed function $g(\cdot)$ for all $k$, we show via simulations in Appendix \ref{sec:appendix_sims} that if $f_k(d_i)$ takes the form in Equation \eqref{eq:fk}, varying $p$ across subpopulations as $p_k$ introduces only relatively minor bias.

\begin{figure}[!t]
    \centering
    \includegraphics[width=\textwidth]{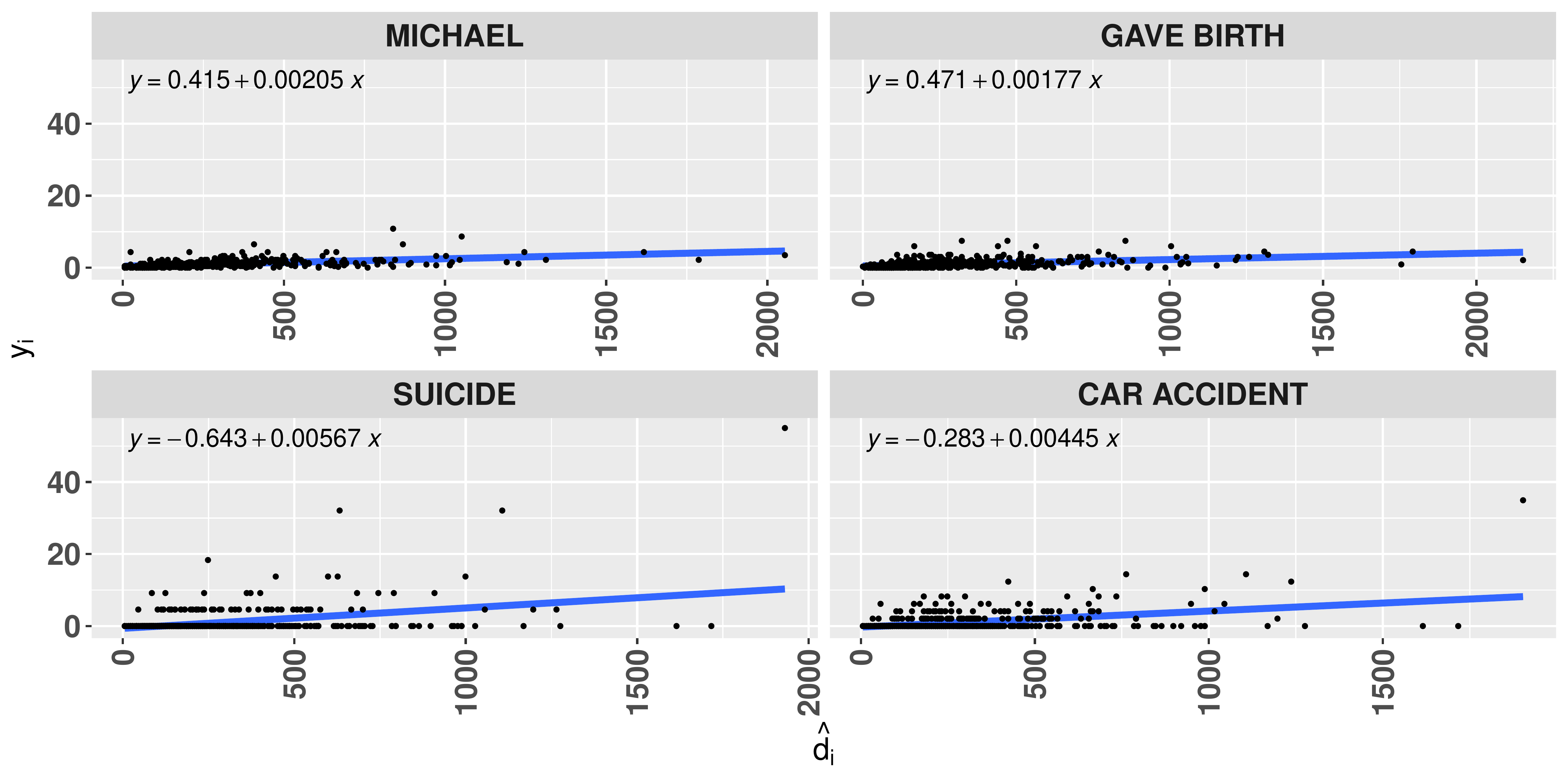}
    \caption{Plot of ARD responses from the McCarty survey against estimated degrees for people who are named Michael, gave birth, committed suicide, or were in a car accident. Estimated linear regression is overlaid.}
    \label{fig:mccarty_degree}
\end{figure}

Our approach to estimate $\gamma_0$ and $\gamma_1$ is detailed in Algorithm \ref{alg:adj}. We motivate the adjustment procedure here using two regression steps. The first estimates the slope between the scaled responses, $y_{ik}/(\frac{1}{n}\sum_{j=1}^n y_{jk})$, and the estimated degrees, $\hat{d}_{i,-k}$. An illustrative example for a real ARD survey is shown in Figure \ref{fig:mccarty_degree}. Larger slopes correspond to subpopulations where respondents with larger degrees report knowing more individuals relative to their own network size than respondents with smaller degrees. Thus, these slopes can be estimated for all subpopulations, including the unknown subpopulation, and measure the relationship between respondent degree and the bias $f_k(d_i)$. We call these slopes ``first-stage slopes.''

Next, we treat these first-stage slopes as covariates in another regression model to model $N_k / \hat{N}_k^{LOO}$ for the known subpopulations and estimate $\gamma_0$ and $\gamma_1$, where $\hat{N}_k^{LOO}$ is the leave-one-out subpopulation size estimate for subpopulation $k$. An illustrative example showing the linear relationship between the observed $N_k / \hat{N}_k^{LOO}$ for the known subpopulations and the first-stage slopes is shown in Figure \ref{fig:sim_slopes} for the two simulation studies in Section \ref{sec:simulation}. Using this relationship, we can then predict $N_H / \hat{N}_H^{LOO}$ for the unknown subpopulation $H$ using the first-stage slope corresponding to $H$. Finally, we can use our predicted $N_H / \hat{N}_H^{LOO}$ to adjust the unknown subpopulation size estimate.

Given the sociological interest in the degree ratio for different subpopulations, we recommend estimating $\delta_k$ for all subpopulations included in the ARD survey, including the subpopulations with known size. One example of an interesting degree ratio corresponds to the ``priest'' subpopulation in the \cite{rwanda2012estimating} NSUM study. \cite{mccarty2001comparing} found that ARD surveys given to clergy yielded larger average network sizes than ARD surveys given to a representative sample. Despite having a known subpopulation size for priests, we are able to observe through the ARD and our proposed approach that priests have substantially larger than average social network sizes.

\newpage
\begin{algorithm}[H]
\SetAlgoLined
\KwResult{Adjusted $N_H$ estimates}
    Set $L$ equal to the number of subpopulations with known sizes $N_k$\;
    Set $K$ equal to the total number of subpopulations\;
    Estimate $\hat{d}_i = N \dfrac{\sum_{k \in known} y_{ik}}{\sum_{k \in known} N_k}$ for all respondents $i$ using all known subpopulations\;
    \For{each k in 1:$L$}{
        Estimate leave-one-out degrees $\hat{d}_{i,-k} = N \dfrac{\sum_{j \in known, j \neq k} y_{ik}}{\sum_{j \in known, j \neq k} N_k}$\;
        Estimate leave-one-out subpopulation sizes, $\hat{N}_k^{LOO} = N \dfrac{\sum_{i = 1}^{n} y_{ik}}{\sum_{i=1}^n \hat{d}_{i,-k}}$\;
        Estimate $\beta_{0,k}$ and $\beta_{1,k}$ for the linear model $\frac{y_{ik}}{\frac{1}{n}\sum_{j=1}^n y_{jk}} = \beta_{0,k} + \beta_{1,k} \hat{d}_{i,-k} + \varepsilon_{i,k}$, $\varepsilon_{i,k} \overset{iid}{\sim} N(0, \sigma_k^2)$\;
    }
    Estimate hidden subpopulation size, $\hat{N}_H = N \dfrac{\sum_{i = 1}^{n} y_{iH}}{\sum_{i=1}^n \hat{d}_i}$\;
    Estimate $\beta_{0,H}$ and $\beta_{1,H}$ for the linear model $\frac{y_{iH}}{\frac{1}{n}\sum_{j=1}^n y_{jH}} = \beta_{0,H} + \beta_{1,H} \hat{d}_{i} + \varepsilon_{i,H}$, $\varepsilon_{i,H} \overset{iid}{\sim} N(0, \sigma_H^2)$\;
    Estimate $\gamma_0$ and $\gamma_1$ for the linear model $N_k / \hat{N}_k^{LOO} = \gamma_0 + \gamma_1 \hat{\beta}_{1,k} + \upsilon_k$, $\upsilon_k \overset{iid}{\sim} N(0, \sigma^2)$, for $k = 1, 2, \ldots, L$\;
    Predict $\hat{\delta}_H = \hat{N}_H / N_H = 1 / (\hat{\gamma}_0 + \hat{\gamma}_1 \hat{\beta}_{1,H})$\;
    Adjust unknown subpopulation size estimate, $\hat{N}_H^{adj} = \hat{N}_H / \hat{\delta}_H$\;
\caption{Degree ratio adjustment procedure}
\label{alg:adj}
\end{algorithm}
\newpage

While we propose the above methodology as a general approach to correct for the degree ratio, the results are based on an assumed form of the data generating process and motivated through empirical results. In practice we recommend using caution when adjusting the size estimates for subpopulations corresponding to names like ``Michael'' or ``Kristina'' in the \cite{mccarty2001comparing} ARD survey. While the popularity of certain names may be related to age and similar demographics, we find that this is dataset dependent and empirically the association is often less pronounced. Applying the degree ratio correction in settings with weak associations risks correcting for spurious relationships in the data rather than for true signals.

\section{Simulation Study}
\label{sec:simulation}

\subsection{Binomial Model}
We simulate ARD from the biased binomial model presented in Equations \eqref{eq:binomial_bias} and \eqref{eq:fk}. We let $p = 2$, although the choice of $p$ does not substantially change the results. To both provide consistent estimates of $\hat{d}_{i,-k}$ and provide a complete range of values, we set the number of respondents at $10000$ and the number of subpopulations at $50$. We set $N = 10^7$ and simulate the subpopulation sizes as $N_k \sim \text{Unif}(10^3, 10^6)$, which corresponds to subpopulation sizes between $0.01\%$ and $10\%$ of the total population size. We simulate the degrees initially as $d_i \sim \text{Unif}(10, 1000)$ and then round them to the nearest integer. Finally, we let $c_k$ range evenly from the minimum possible value to the maximum possible value, such that the binomial probability is between 0 and 1 for all respondents. For this simulation, $c_k$ ranges from $-1.5 \times 10^{-6}$ to $1.5 \times 10^{-6}$, resulting in $\hat{N}_k / N_k$ ranging from $0.82$ to $1.37$. We implement Algorithm \ref{alg:adj} in a leave-one-out procedure, where we estimate the adjusted subpopulation sizes for each subpopulation sequentially, treating each successive subpopulation as unknown.

We plot the estimated first-stage slopes against $N_k / \hat{N}_k$ for the 50 subpopulations in Figure \ref{fig:binomial_slope}, showing that the estimated slopes are almost perfectly correlated with $N_k / \hat{N}_k$. The point farthest from the best-fit line corresponds to the smallest subpopulation with prevalence $0.02\%$, which is relatively small for subpopulations included in NSUM surveys. Despite the small size, the departure from the expected value is minimal.

The performance of the adjusted estimator is shown in Figure \ref{fig:binomial_results} and Table \ref{tab:mape}. The reduction in mean absolute percent error is 97\%, indicating that we almost perfectly recover the true size estimates. Most of the remaining error is from the smallest subpopulation above. This simulation shows that despite estimating $\hat{d}$ and $\hat{d}_{i,-k}$, we are still able to adjust for most of the subpopulation size error introduced by the degree ratio.

Appendix \ref{sec:appendix_sims} contains an additional simulation based on the binomial model, but where $p$ is allowed to vary across subpopulations. The results indicate that the correlation between $\hat{N}_k / N_k$ and the first-stage slopes are still fairly linear, although with larger variation. These results indicate that our proposed approach will likely work for real data where the behavior may differ considerably between subpopulations.

\subsection{Stochastic Block Model}
We simulate a network from a stochastic block model (SBM) with 20000 respondents and 20 groups. We set each group size to be 1000. In order to have a range of connectivity, the within-group connectivities (i.e. the diagonal of the connectivity matrix) are given by a sequence from 0.25 to 0.5 in steps of 0.05. All between-group connectivity probabilities are 0.05. These parameters were chosen to provide a sufficient sample size to generate ARD with realistic values and to provide a range of degree ratios across subpopulations of equal sizes.

In order to evaluate the model performance, we again implemented Algorithm \ref{alg:adj}, but using the true respondent degrees directly rather than estimating them. This is necessary under this simulation design because estimates vary widely for each leave-one-out step, unlike for more traditional ARD data where leave-one-out degree estimates are relatively stable. We perform the same leave-one-out procedure as for the binomial model simulation study.

\begin{figure}[!t]
    \centering
    \begin{subfigure}{0.7\textwidth}
        \includegraphics[width=\textwidth]{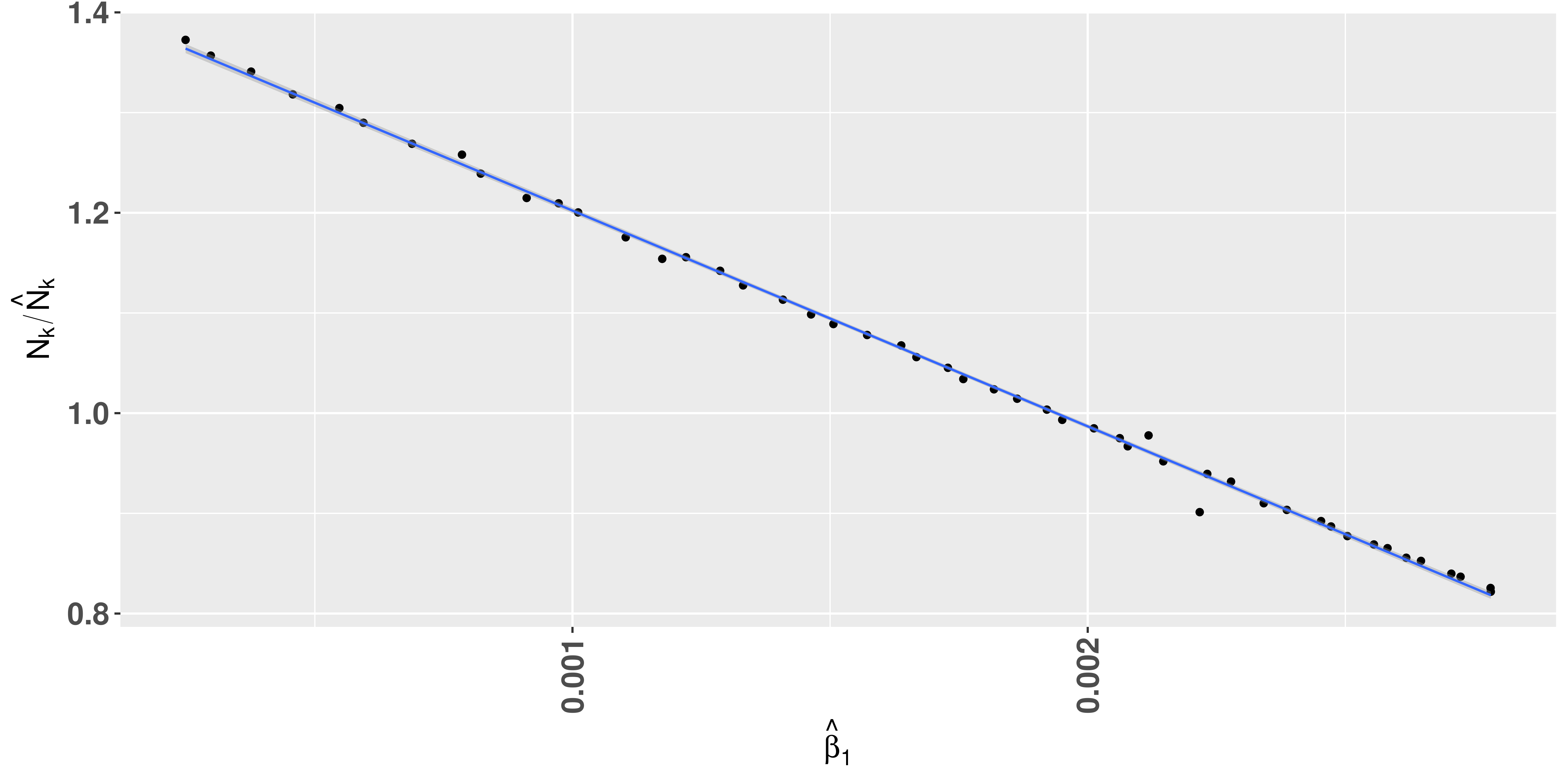}
        \caption{}
        \label{fig:binomial_slope}
    \end{subfigure}\\
    \begin{subfigure}{0.7\textwidth}
        \includegraphics[width=\textwidth]{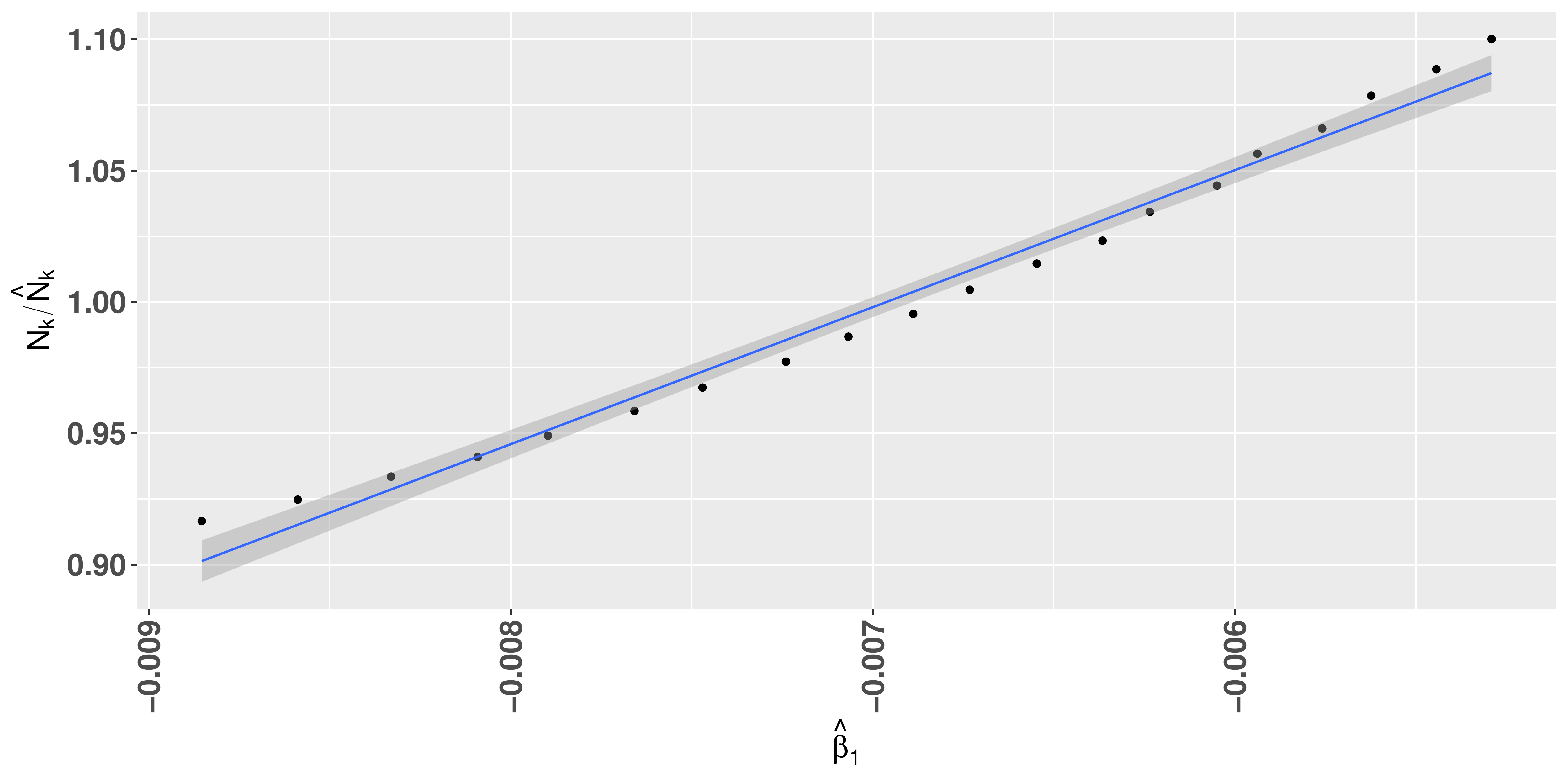}
        \caption{}
        \label{fig:sbm_slope}
    \end{subfigure}
    \caption{Empirical error of $N_k / \hat{N}_k$ plotted against the estimated first-stage slopes for the binomial model simulation (a) and for the stochastic block model simulation (b).}
    \label{fig:sim_slopes}
\end{figure}

We plot the estimated first-stage slopes against $N_k / \hat{N}_k$ for the 20 subpopulations in Figure \ref{fig:sbm_slope}, again showing that the estimated slopes are highly correlated with $N_k / \hat{N}_k$, although the relationship is slightly non-linear, unlike for the binomial model.

\begin{figure}[!t]
    \centering
    \begin{subfigure}{0.7\textwidth}
        \includegraphics[width=\textwidth]{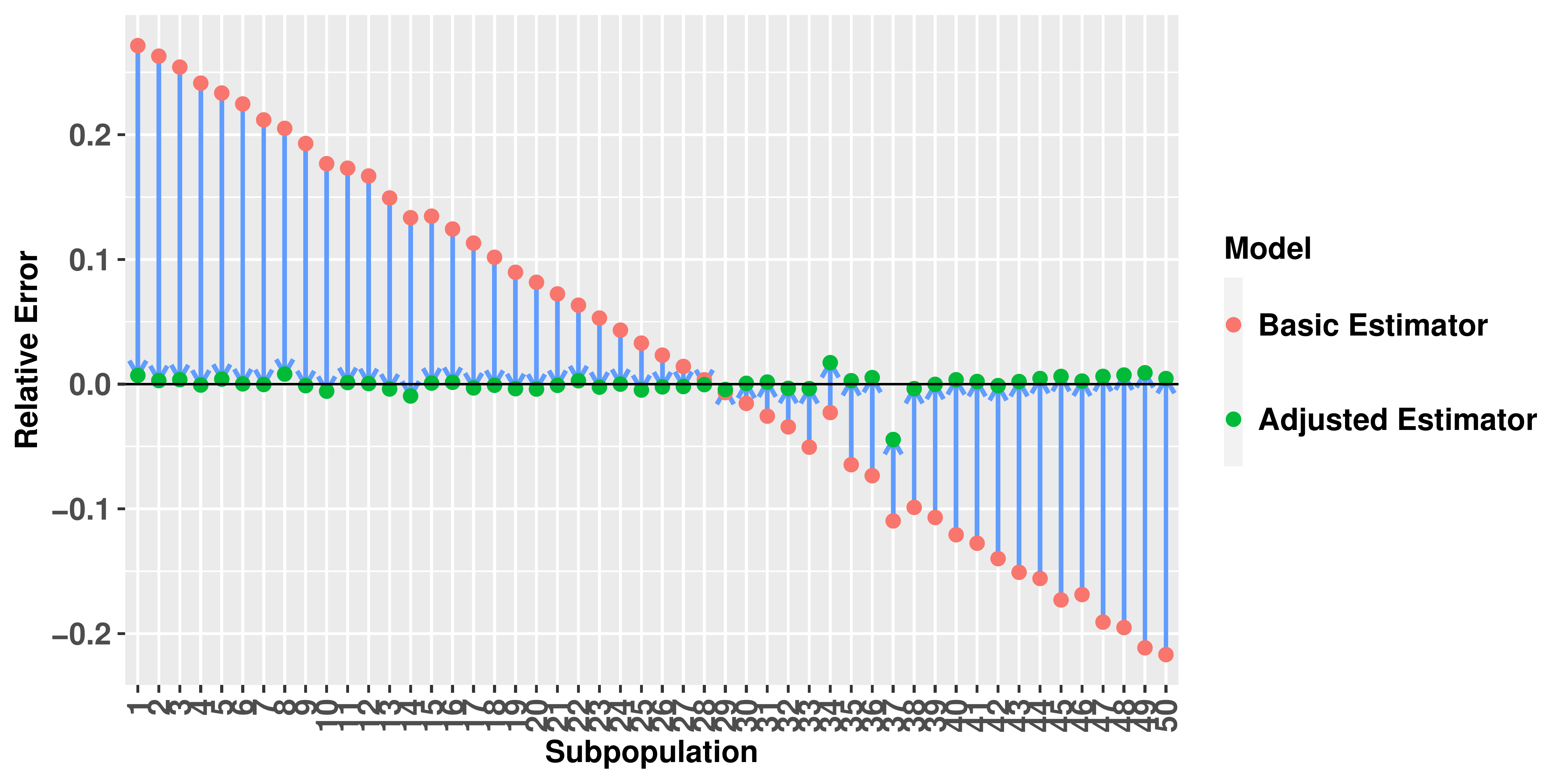}
        \caption{}
        \label{fig:binomial_results}
    \end{subfigure}\\
    \begin{subfigure}{0.7\textwidth}
        \includegraphics[width=\textwidth]{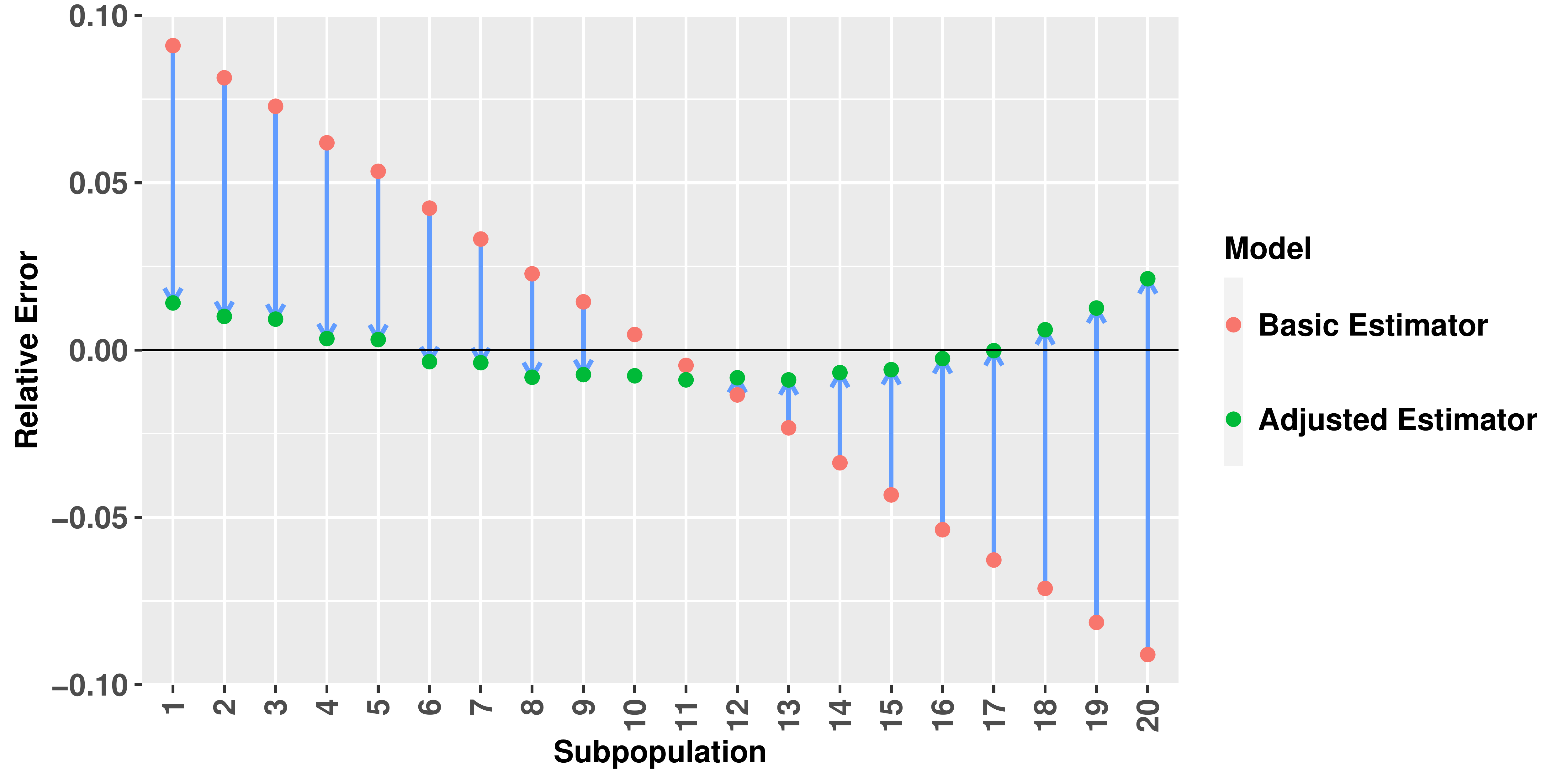}
        \caption{}
        \label{fig:sbm_results}
    \end{subfigure}
    \caption{Relative error subpopulation size estimates for the binomial model (a) and for the stochastic block model simulation (d). Original basic scale-up estimator and adjusted basic scale-up estimator estimates are shown in pink and green, respectively. Relative error is calculated by $100*(Truth - estimates) / Truth$. Subpopulations are ordered from smallest to largest. Arrows indicate subpopulations where the adjusted estimates have smaller relative errors.}
    \label{fig:sim_results}
\end{figure}

The results for this simulation study are shown in Figure \ref{fig:sbm_results}, where the original basic scale-up estimator estimates are shown in pink, our adjusted estimates in green, and blue arrows indicate subpopulations where our adjusted estimates have smaller absolute relative error. In this study, we outperform the basic scale-up estimator for all 20 subpopulations. The percent reduction in mean absolute percent error is presented in Table \ref{tab:mape}. For this simulation study, adjusting the size estimates resulted in a $84\%$ reduction in mean absolute percent error. It is clear in this simulation study that our proposed method is able to consistently correct for the degree ratio and substantially improve the existing basic scale-up estimator estimates despite the data coming from a model different than our assumed binomial likelihood.

\section{Network Scale-up Method Studies}
\label{sec:realdata}

In this section, we apply our adjustment procedure to two real ARD surveys. We show that despite its simplicity, the proposed adjustment substantially improves size estimates. We follow the same procedure outlined in the simulation study to evaluate the performance of our proposed methods, where we again estimate $\hat{d}_{i,-k}$ when studying subpopulation $k$. This matches the procedure used when estimating the unknown subpopulations, where only the ARD responses in the known subpopulations are used to estimate the degrees and subpopulation sizes.

\subsection{McCarty ARD Study}

First, we apply our proposed adjustment method to the ARD first collected and presented in \cite{mccarty2001comparing}. This dataset contains responses from 574 respondents about 32 subpopulations, 3 of which are unknown (individuals who are homeless, have been raped, or are HIV positive). Twelve of the 29 known subpopulations corresponds to names. We remove 53 respondents for having 1 or more missing responses (47 of those 53 respondents had only 1 missing response), resulting in 521 respondents. As the primary purpose of this work is to evaluate the performance of our proposed adjusted estimator compared to the basic scale-up estimator, we do not study the effect of removing these respondents with missing data.

\begin{figure}[!t]
    \centering
    \begin{subfigure}{0.8\textwidth}
        \includegraphics[width=\textwidth]{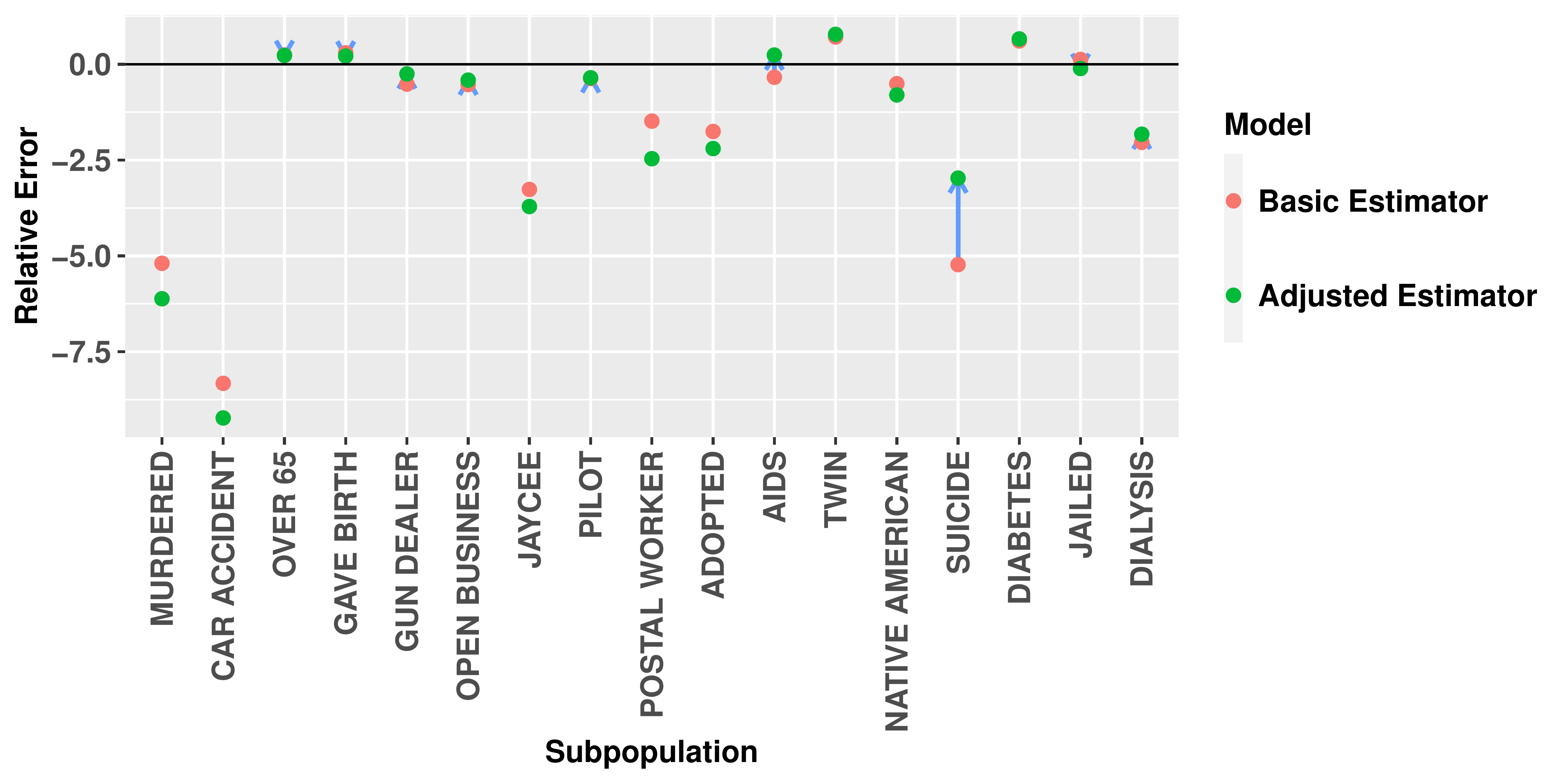}
        \caption{}
        \label{fig:mccarty_non_names}
    \end{subfigure}
        \begin{subfigure}{0.8\textwidth}
        \includegraphics[width=\textwidth]{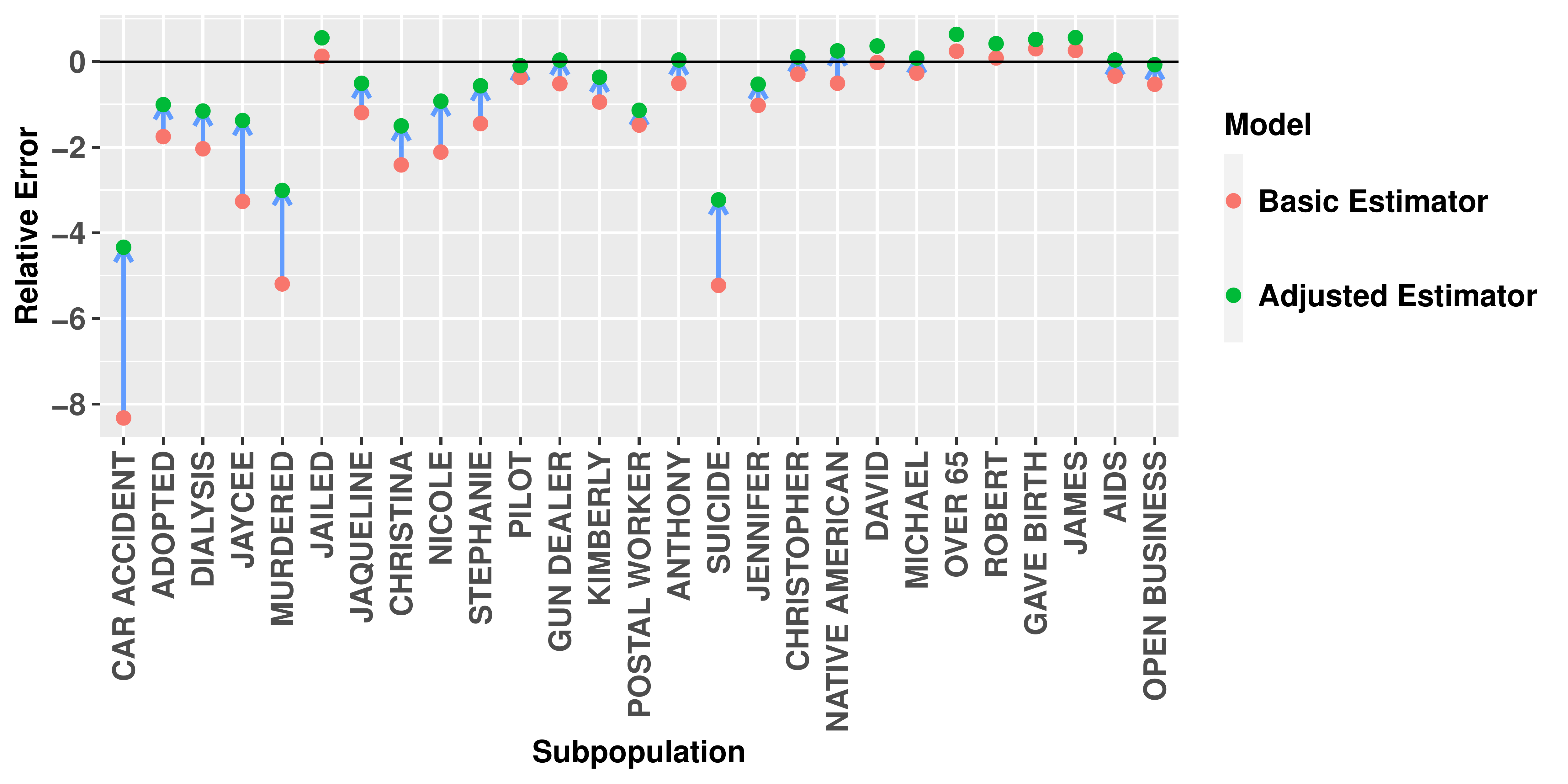}
        \caption{}
        \label{fig:mccarty_twin_results}
    \end{subfigure}
    \caption{Relative error subpopulation size estimates for non-name (a), and for the non-twin and non-diabetes subpopulations of the McCarty study (b). Original basic scale-up estimator and adjusted basic scale-up estimator estimates are shown in pink and green, respectively. Relative error is calculated by $100*(Truth - estimates) / Truth$. Subpopulations are ordered from smallest to largest. Arrows indicate subpopulations where the adjusted estimates have smaller relative errors.}
    \label{fig:mccarty_real_results}
\end{figure}

The percent reduction in mean absolute percent error for different subsets of subpopulations are shown in Table \ref{tab:mape}. For this dataset, substantial improvements exist when adjusting subpopulations corresponding to names, where based on Figure \ref{fig:mccarty_slopes}, there seems to be a strong linear relationship between $N_k / \hat{N}_k$ and the first-stage slopes for the names, but a weaker relationship with higher variance for the non-name subpopulations. One potential explanation of this results is that the adjustment depends heavily on the bias introduced by the ``twin'' and ``diabetes'' subpopulations, which likely have high transmission effect, but relatively low barrier effects. Practitioners may choose to remove these two subpopulations from the second-stage regression based on additional information about these populations, thus improving the performance of the adjustment procedure. We compare the final adjusted size estimates for the 17 non-name subpopulations against the original basic scale-up estimator size estimates in Figure \ref{fig:mccarty_non_names}. The adjusted estimator results in a 36\% reduction in mean absolute percent error. The results after keeping the name groups but removing the twin and diabetes groups are shown in Figure \ref{fig:mccarty_twin_results} and resulted in a $43\%$ reduction in mean absolute percent error.

\subsection{Rwanda Meal ARD Study}

Next, we consider the Rwanda Meal ARD survey \citep{rwanda2012estimating, feehan2016quantity}. In 2011, researchers collected ARD from 4,669 respondents in Rwanda in order to estimate the size of four key populations: female sex workers (FSW), male clients of sex workers (MCSW), men who have sex with men (MSM), and people who inject drugs (IDU). Thirteen of the 22 known subpopulations correspond to names. Rwanda Biomedical Center/Institute of HIV/AIDS, Disease Prevention and Control (RBC/IHDPC) and their partners require accurate size estimates of these unknown subpopulations in order to plan and implement efficient HIV prevention strategies for current HIV cases and understand the trend of HIV cases across time.

One of the primary motivations of the survey was to compare the results of NSUM size estimates between two definitions of whether a respondent ``knows'' someone \citep{feehan2016quantity}. The first definition, called the \textit{acquaintance} definition, quantifies the ``people the respondent has had some contact with --- either in person, over the phone, or on the computer in the previous 12 months." The \textit{meal} definition restricts the acquaintance definition, quantifying the ``people the respondent has shared a meal or drink with in the past 12 months, including family members, friends, coworkers, or neighbors, as well as meals or drinks taken at any location, such as at home, at work, or in a restaurant." \cite{feehan2016quantity} were able to show that estimates from the meal definition were consistently closer to the known sizes than estimates from the acquaintance definition. While the authors were unable to confidently extend this finding to subpopulations with unknown size (e.g. FSW), it is not unlikely that these estimates for unknown subpopulations would also be more accurate.

In order to use the dataset least prone to errors, for our analysis, we consider only the dataset collected from the meal definition. Given that the meal definition implies a stronger relationship between the respondent and their social connections, it is reasonable to assume that the respondent knows more about each person they recalled, reducing the transmission error. Furthermore, given that the pool of potential connections is smaller, respondents should have an easier time recalling everyone in a given subpopulation, also reducing recall error. In order to show that our proposed method accurately accounts for the bias introduced by differences in average network sizes between groups, it is helpful to use a dataset that faces smaller biases from other sources.

In this study, we analyzed responses from 2405 respondents about 22 known subpopulations. Only one respondent was removed for a missing response to how many people they know who are Muslims.

\begin{figure}[!t]
    \centering
        \centering
    \begin{subfigure}{0.8\textwidth}
        \includegraphics[width=\textwidth]{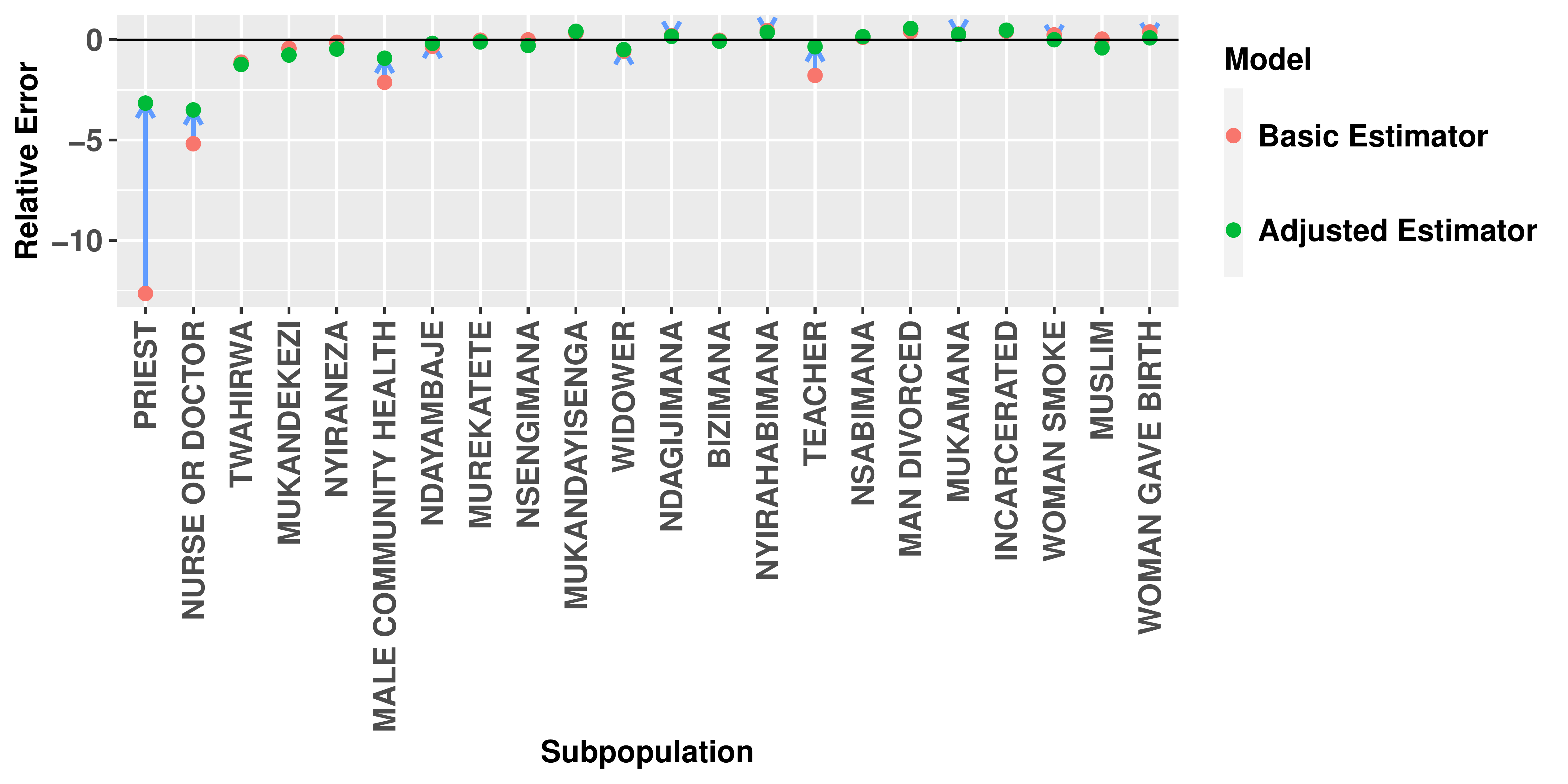}
        \caption{}
        \label{fig:rwanda_all}
    \end{subfigure}
        \begin{subfigure}{0.8\textwidth}
        \includegraphics[width=\textwidth]{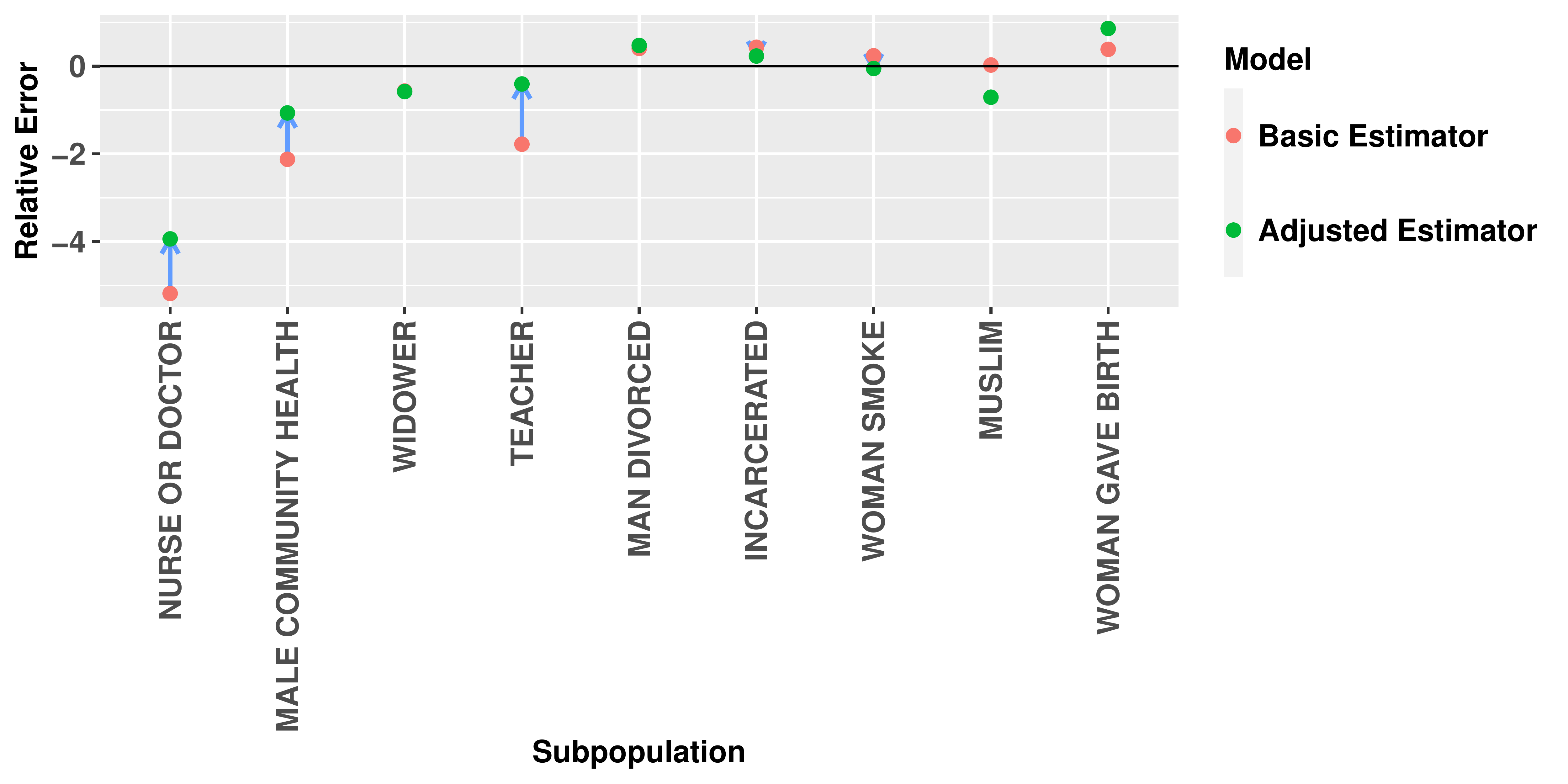}
        \caption{}
        \label{fig:rwanda_non_names}
    \end{subfigure}
    \caption{Relative error subpopulation size estimates for all subpopulations (a), and for the non-name and non-priest subpopulations of the Rwanda Meal study (b). Original basic scale-up estimator and adjusted basic scale-up estimator estimates are shown in pink and green, respectively. Relative error is calculated by $100*(Truth - estimates) / Truth$. Subpopulations are ordered from smallest to largest. Arrows indicate subpopulations where the adjusted estimates have smaller relative errors.}
    \label{fig:rwanda_real_results}
\end{figure}

The percent reduction in mean absolute percent error for different subsets of subpopulations are again shown in Table \ref{tab:mape}. We compare the relative error of the basic scale-up estimator and our adjusted estimates in Figure \ref{fig:rwanda_all} for all known subpopulations and in Figure \ref{fig:rwanda_non_names} for non-name and non-priest known subpopulations. When considering all subpopulations, the adjusted estimator has a 47\% reduction in mean absolute percent error. While our adjusted estimator performs best when including the priest group, we remove this group from Figure \ref{fig:rwanda_non_names} to show that even after removing highly influential groups like priest, our adjusted estimator still outperforms the basic scale-up estimator. Visually, adjusting the estimate via our approach substantially improves the overall performance of the basic scale-up estimator. Numerically, our adjusted estimates perform better in four of the nine non-name and non-priest known subpopulations. However, the adjusted estimators perform significantly better than the basic scale-up estimator for those four subpopulations, while only performing slightly worse for the remaining subpopulations. Our adjusted estimator reduced the mean absolute percent error 25\% for the non-name and non-priest groups. With the priest subpopulation included, the percent reduction in mean absolute percent error is reduced by 64\%. Using all known groups, the reduction is 47\%. The adjusted size estimates are substantially better when including the priest subpopulation because priests have relatively large social networks and have significantly larger social networks than other members of the population, emphasizing that our proposed methods works especially well when there are clear differences in social network sizes across subpopulations \citep{mccarty2001comparing}.

\begin{figure}[!t]
    \centering
    \begin{subfigure}{0.8\textwidth}
        \includegraphics[width=\textwidth]{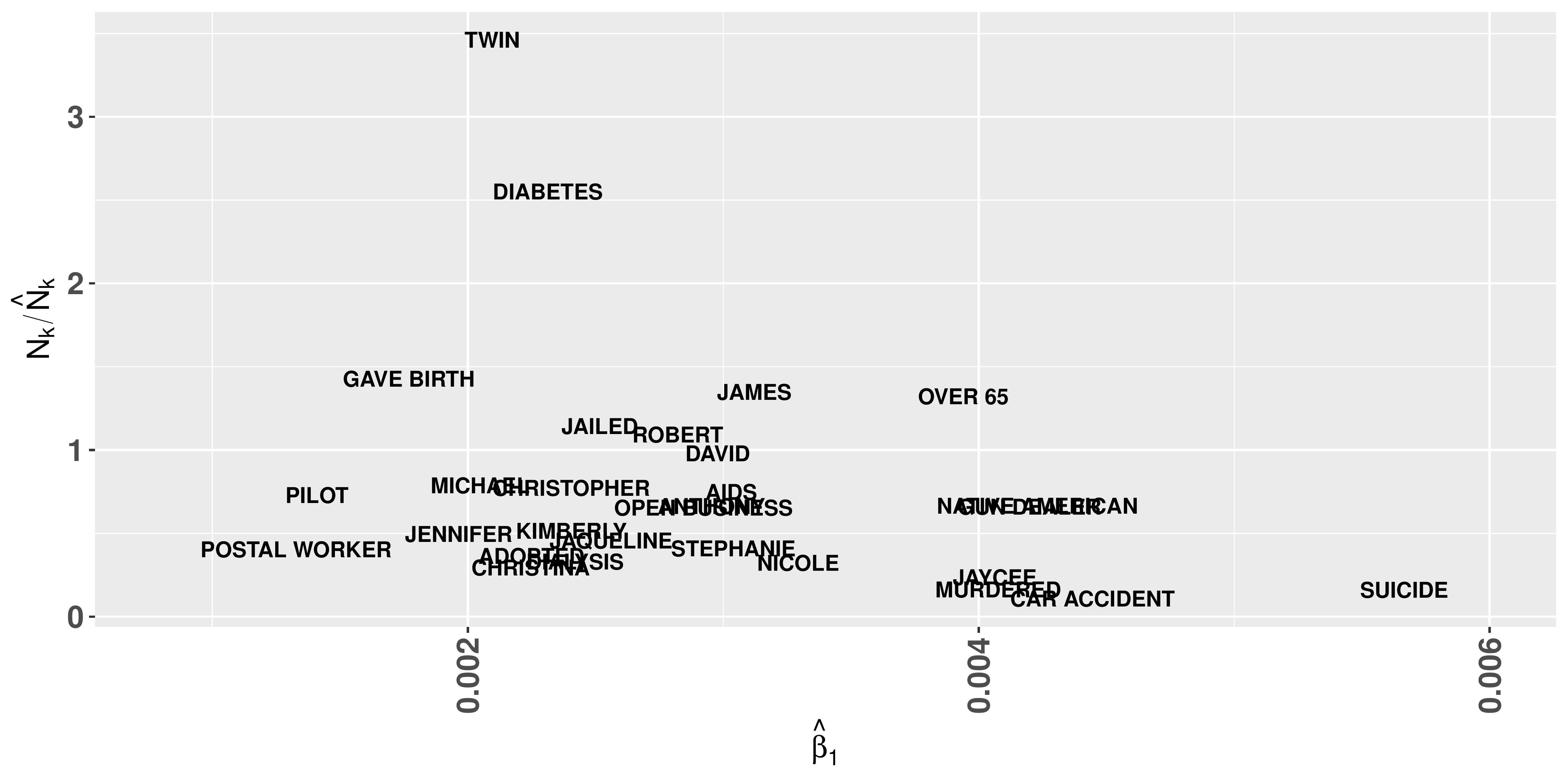}
        \caption{}
        \label{fig:mccarty_slopes}
    \end{subfigure}
        \begin{subfigure}{0.8\textwidth}
        \includegraphics[width=\textwidth]{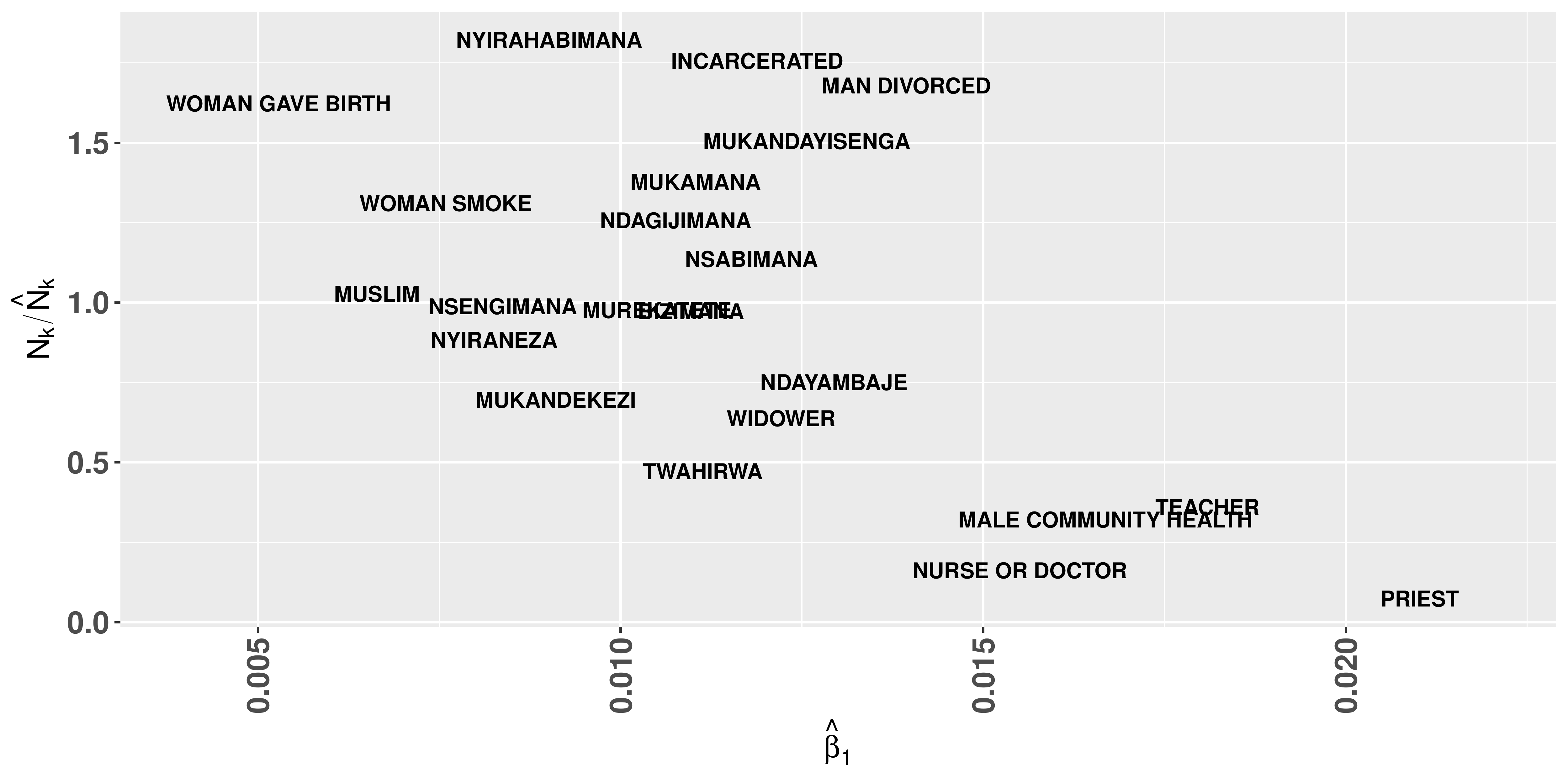}
        \caption{}
        \label{fig:rwanda_slopes}
    \end{subfigure}
    \caption{Empirical error of $N_k / \hat{N}_k$ plotted against the estimated first-stage slopes for the McCarty study (a) and Rwanda Meal study (b).}
    \label{fig:real_slopes}
\end{figure}

Unlike for the \cite{mccarty2001comparing} dataset, the proposed adjustment does not work well for name-based groups. Adjusting size estimates for only the name-based groups results in a 12\% decrease in mean absolute percent error. Based on Figure \ref{fig:rwanda_slopes}, we see that for the name groups, $N_k / \hat{N}_k$ is not highly correlated with the first-stage slopes.

\FloatBarrier

\section{Discussion}
\label{sec:improve_discussion}

We have demonstrated through both simulations and through two data examples that our proposed degree ratio adjustment can substantially reduce the bias of the basic scale-up estimator. \cite{mcpherson2001birds} found that homophily of social networks exists for a variety of groups, including those characterized by behaviors, attitudes, and occupations. This observations lends some credibility towards the assumed form of the bias term $f_k(d_i)$ for both the known and unknown subpopulations since different groups may form social networks in similar ways. We rely on the performance of our proposed approach with respect to the known subpopulations out of necessity since we are unable to access the quality of NSUM estimators for the unknown subpopulations.

The key novelty of this paper is that our proposed method handles the very difficult problem of varying average network sizes across different subpopulations \textit{without} using auxiliary data. Methods that use auxiliary data may intuitively perform better than our combined procedure and we encourage researchers to use additional data when available. However, collecting additional data is often impossible, necessitating an approach that recycles the available data.

An interesting direction for future work is to consider how the proposed degree ratio adjustment affects the choice of known subpopulations in the NSUM survey. Previous researchers have proposed relying heavily on known groups corresponding to names, since these groups may be subject to fewer and smaller biases. However, if the adjustment procedure relies on estimating the relationship between the first-stage slopes and the estimator bias, it may actually be advantageous to instead include groups in the survey that provide a wider range of estimator bias than name-based groups since this may lead to more accurate adjustments.

Furthermore, as we presented in this work, NSUM models should be evaluated using performance metrics that do not favor large subpopulations. Metrics like root mean squared error are dominated by these large populations like ``people who have diabetes'' or ``people who are twins'' so that the accuracy of size estimates corresponding to populations like ``people who were murdered'' or ``people who committed suicide'' are not influential.

As with all methods used to estimate the size of hard-to-reach subpopulations, it is difficult to understand, model, and account for all sources of bias. In some cases, accounting for one source of bias may result in worse estimates if the other sources of bias are ignored. Continued research is needed to understand how the different NSUM biases interact together and whether it suffices to account for each form of bias independently. We believe the NSUM holds an important role in providing accurate, quick, and affordable size estimates and urge future researchers to continue developing this promising method.

\subsection*{Data Availability Statement}
The datasets analyzed during the current study are publicly available, but we do not have permission to distribute them. All code used to create the results presented in this manuscript are available at \url{https://github.com/XXXX}.

\newpage

\section{Tables and Figures}
\FloatBarrier

\begin{table}[!h]
\centering
\caption{Percent reduction in mean absolute percent error (MAPE) for the adjusted size estimates for the SBM simulation, McCarty, and Rwanda Meal studies. Percent reduction is calculated by $100 * (MAPE^{basic} - MAPE^{adjusted}) / (MAPE^{basic})$.}
\begin{tabular}{|l|l|c|}
\hline
\textbf{Data Set} & \textbf{Subpopulations} & \textbf{Adjusted} \\ \hline
\textbf{Binomial Simulation} & All & 97\% \\ \hline
\textbf{SBM Simulation} & All & 84\% \\ \hline
\multirow{3}{*}{\textbf{McCarty}} & All & 36\% \\ \cline{2-3} 
 & Non-names & -3\% \\ \cline{2-3} 
 & All but twin/diabetes & 43\% \\ \cline{2-3} 
 & Names & 33\% \\ \hline
 \multirow{3}{*}{\textbf{Rwanda Meal}} & All & 47\% \\ \cline{2-3} 
 & All but priest & 15\% \\ \cline{2-3} 
 & Non-names, with priest & 64\% \\ \cline{2-3} 
 & Non-names, no priest & 25\% \\ \cline{2-3}
 & Names & -12\% \\ \cline{2-3} 
 \hline
\end{tabular}%
\label{tab:mape}
\end{table}

\newpage

\FloatBarrier

\FloatBarrier

\FloatBarrier

\newpage

\appendix
\appendixpage
\section{Proofs}
\label{sec:proofs}

\subsection{Proof of Proposition 2}

\begin{proof}
Given perfect link reporting, the numerator remains unchanged from Proposition 1, where
\begin{equation*}
    E\left[\sum_{i\in S} (y_{iH}/\pi_i)\right] = \sum_{i=1}^N y_{iH} = \sum_{j \in H} d_j = \text{sum of degrees in subpopulation H}.
\end{equation*}
For the denominator, we have 
\begin{align*}
    E\left\{\frac{1}{N} \sum_{i\in S} \left[\left(\sum_{k=1}^L y_{ik} / \sum_{k=1}^L N_k \right)/\pi_i \right]\right\} &= \frac{1}{N \sum_{k=1}^L N_k} E\left(\sum_{i\in S} \sum_{k=1}^L y_{ik} / \pi_i \right) \\
    &= \frac{1}{N \sum_{k=1}^L N_k} E\left(\sum_{k=1}^L \sum_{i\in S} y_{ik} / \pi_i \right) \\
    &= \frac{1}{N \sum_{k=1}^L N_k}\left(\sum_{k=1}^L \sum_{i = 1}^N y_{ik} \right) \\
    &= \frac{1}{N \sum_{k=1}^L N_k}\left(\sum_{k=1}^L \sum_{j \in k} d_j \right) \\
    &= \frac{1}{N \sum_{k=1}^L N_k}\left(\sum_{k=1}^L N_k \bar{d}_k \right)
\end{align*}
Then, since the ratio of two unbiased estimators is approximately unbiased \citep{sarndal2003model}, the expected value of the estimator in Equation (\ref{eq:mle_unknown}) is approximately given by
\begin{equation*}
    E\left[ \hat{N}_H \right] \approx \frac{N_H \bar{d}_H \sum_{k=1}^L N_k}{ \sum_{k=1}^L \bar{d}_k N_k },
\end{equation*}
where $\bar{d}_k$ represents the average degree of members of subpopulation $k$.
\end{proof}

\section{Proof of Proposition 3}
\label{sec:proof3}

\begin{proof}
Let
\begin{equation}
    y_{ik} \sim Binomial\left(d_i, \frac{N_k}{N} f_k(d_i)\right),
\end{equation}
for $i = 1, \ldots, n$ and $k = 1, \ldots, L$, where
\begin{equation}
    f_k(d_i) = 1 + g(d_i) c_k
\end{equation}
and where the $d_i$ and $c_k$ are known. We prove that $E\left(N_k/\hat{N}_k\right) \approx E(\hat{r}_k)$, where
\begin{equation*}
    \frac{N_k}{\hat{N}_k} = \frac{N_k \sum_{i=1}^n d_i}{N \sum_{i=1}^n y_{ik}},
\end{equation*}
and
\begin{equation*}
    \hat{r}_k = \gamma_0 + \gamma_1 \left( \frac{\sum_{i=1}^n (d_i - \bar{d}_i)z_{ik}}{\sum_{i=1}^n (d_i - \bar{d}_i)^2} \right),
\end{equation*}
where $z_{ik}$ are calculated by scaling the responses by their column means, i.e.
\begin{equation*}
    z_{ik} = \frac{y_{ik}}{\frac{1}{n}\sum_{j=1}^n y_{jk}}.
\end{equation*}

First, we find $E\left(N_k/\hat{N}_k\right)$. Since the ratio of two unbiased estimators is approximately unbiased \citep{sarndal2003model},
\begin{align*}
    E\left(\frac{N_k \sum_{i=1}^n d_i}{N \sum_{i=1}^n y_{ik}} \right) &\approx \frac{E\left(N_k \sum_{i=1}^n d_i\right)}{E\left(N \sum_{i=1}^n y_{ik}\right)}\\
    &=\frac{N_k \sum_{i=1}^n d_i}{N \sum_{i=1}^n E(y_{ik})} \\
    &= \frac{N_k \sum_{i=1}^n d_i}{N \sum_{i=1}^n (d_i N_k f_k(d_i)/N)} \\
    &= \frac{\sum_{i=1}^n d_i}{\sum_{i=1}^n d_i f_k(d_i)}.
\end{align*}

Second, we find $E(\hat{r}_k)$. Note, we can write $\hat{r}_k$ as a function of the first-stage regression slopes, given by
\begin{equation*}
    \hat{r}_k = \gamma_0 + \gamma_1 \hat{\beta}_k, \qquad \hat{\beta}_k = \frac{\sum_{i=1}^n (d_i - \bar{d}_i)z_{ik}}{\sum_{i=1}^n (d_i - \bar{d}_i)^2}, \quad k = 1, \ldots, L.
\end{equation*}
Thus, $E(\hat{r}_k) = \gamma_0 + \gamma_1 E(\hat{\beta}_k)$. The expectation of $z_{ik}$ is approximately given by
\begin{align*}
    E(z_{ik}) &= E\left(\frac{1}{\frac{1}{n} \sum_{j=1}^n y_{jk} / y_{ik}} \right) \\
    &\approx \frac{1}{\frac{1}{n} \sum_{j=1}^n E\left(y_{jk} / y_{ik} \right)}.
\end{align*}
For $i = j$,
\begin{equation*}
    E\left(y_{jk} / y_{ik} \right) = 1.
\end{equation*}
For $i \neq j$,
\begin{equation*}
    E\left(y_{jk} / y_{ik} \right) \approx \frac{E(y_{jk})}{E(y_{ik})}.
\end{equation*}
Thus,
\begin{align*}
    E(z_{ik}) &\approx \frac{1}{\frac{1}{n} \sum_{j=1}^n \frac{E(y_{jk})}{E(y_{ik})}} \\
    &= \frac{E(y_{ik})}{\frac{1}{n} \sum_{j=1}^n E(y_{jk})} \\
    &= \frac{N_k d_i f_k(d_i)}{\frac{1}{n} \sum_{j=1}^n N_k d_j f_k(d_j)} \\
    &= \frac{d_i  f_k(d_i)}{\frac{1}{n} \sum_{j=1}^n d_j f_k(d_j)},
\end{align*}
and,
\begin{align*}
    E\left(\hat{\beta}_k  \right) &= \frac{\sum_{i=1}^n (d_i - \bar{d}_i)E(z_{ik})}{\sum_{i=1}^n (d_i - \bar{d}_i)^2} \\
    &\approx \frac{\sum_{i=1}^n (d_i - \bar{d}_i)\frac{d_i  f_k(d_i)}{\frac{1}{n}\sum_{j=1}^n d_j f_k(d_j)}}{\sum_{i=1}^n (d_i - \bar{d}_i)^2} \\
    &\propto \frac{\sum_{i=1}^n (d_i - \bar{d}_i) d_i f_k(d_i)}{\sum_{j=1}^n d_j f_k(d_j)} \\
    &\propto \frac{\sum_{i=1}^n d_i^2 f_k(d_i)}{\sum_{j=1}^n d_j f_k(d_j)},
\end{align*}
where the proportionality is with respect to $k$. For a general $f_k(d_i)$, we have
\begin{equation*}
    \frac{\sum_{i=1}^n d_i}{\sum_{i=1}^n d_i f_k(d_i)} \approx \gamma_0 + \gamma_1 \frac{\sum_{i=1}^n d_i^2 f_k(d_i)}{\sum_{i=1}^n d_i f_k(d_i)}
\end{equation*}
Finally, we show $\gamma_0$ and $\gamma_1$ are independent of $k$. Consider two points for $k = 1$ and $k = 2$, without loss of generality. Plugging in our specific form of $f_k(d_i)$, we find $\gamma_1$ by solving
\begin{align*}
    \gamma_1 &= \frac{\frac{\sum_{i=1}^n d_i}{\sum_{i=1}^n d_i (a + g(d_i) c_2)} - \frac{\sum_{i=1}^n d_i}{\sum_{i=1}^n d_i (a + g(d_i) c_1)}}{\frac{\sum_{i=1}^n d_i^2 (a + g(d_i) c_2)}{\sum_{i=1}^n d_i (a + g(d_i) c_2)} - \frac{\sum_{i=1}^n d_i^2 (a + g(d_i) c_1)}{\sum_{i=1}^n d_i (a + g(d_i) c_1)}}  \\ \\
    &= \frac{\frac{\left[\sum_{i=1}^n d_i\right]\left[\sum_{i=1}^n d_i (a + g(d_i) c_1)\right] - \left[\sum_{i=1}^n d_i\right]\left[\sum_{i=1}^n d_i (a + g(d_i) c_2)\right]}{\left[\sum_{i=1}^n d_i (a + g(d_i) c_2)\right]\left[\sum_{i=1}^n d_i (a + g(d_i) c_1)\right]}}{\frac{\left[\sum_{i=1}^n d_i (a + g(d_i) c_2)\right]\left[\sum_{i=1}^n d_i^2 (a + g(d_i) c_1)\right] - \left[\sum_{i=1}^n d_i (a + g(d_i) c_1)\right]\left[\sum_{i=1}^n d_i^2 (a + g(d_i) c_2)\right]}{\left[\sum_{i=1}^n d_i (a + g(d_i) c_2)\right]\left[\sum_{i=1}^n d_i (a + g(d_i) c_1)\right]}} \\ \\
    &= \frac{\left[\sum_{i=1}^n d_i\right]\left[\sum_{i=1}^n d_i (a + g(d_i) c_1)\right] - \left[\sum_{i=1}^n d_i\right]\left[\sum_{i=1}^n d_i (a + g(d_i) c_2)\right]}{\left[\sum_{i=1}^n d_i (a + g(d_i) c_2)\right]\left[\sum_{i=1}^n d_i^2 (a + g(d_i) c_1)\right] - \left[\sum_{i=1}^n d_i (a + g(d_i) c_1)\right]\left[\sum_{i=1}^n d_i^2 (a + g(d_i) c_2)\right]} \\ \\
    &= \frac{(c_1 - c_2) \left[\left(\sum_{i=1}^n d_i\right) \left(\sum_{i=1}^n d_i g(d_i)\right) \right]}{a(c_1 - c_2) \left[ \left(\sum_{i=1}^n d_i\right) \left(\sum_{i=1}^n d_i^2 g(d_i)\right) - \left(\sum_{i=1}^n d^2_i \right) \left(\sum_{i=1}^n d_i g(d_i)\right) \right]} \\ \\
    &= \frac{\left[\left(\sum_{i=1}^n d_i\right) \left(\sum_{i=1}^n d_i g(d_i)\right) \right]}{a\left[ \left(\sum_{i=1}^n d_i\right) \left(\sum_{i=1}^n d_i^2 g(d_i)\right) - \left(\sum_{i=1}^n d^2_i \right) \left(\sum_{i=1}^n d_i g(d_i)\right) \right]}
\end{align*}
From the above, $\gamma_1$ exists and is independent of $k$ for any $a \neq 0$ and any function $g(d_i)$ that does not depend on $k$. Thus,
\begin{equation*}
     E\left(\frac{N_k}{\hat{N}_k}\right) \approx E\left(\gamma_0 + \gamma_1 \left( \frac{\sum_{i=1}^n (d_i - \bar{d}_i)y_{ik}}{\frac{1}{n}\left(\sum_{i=1}^n y_{ik}\right) \left(\sum_{i=1}^n (d_i - \bar{d}_i)^2\right)} \right) \right)
\end{equation*}
for some $\gamma_0$ and $\gamma_1$, which are independent of $k$.
\end{proof}

\section{Additional Results}
\label{sec:appendix_sims}

Here we study the behavior of the adjustment procedure from simulated data for different values of $p$ across the subpopulations. We simulate ARD from the biased binomial model presented in Section \ref{sec:degree_ratio}. We let $p$ vary between $-2$, $-1$, $1$, and $2$ for different subpopulations. We set the number of respondents at $10,000$ and the number of subpopulations at $50$. We set $N = 10^7$ and simulate the subpopulation sizes as $N_k \sim \text{Unif}(10^3, 10^6)$, which corresponds to subpopulation sizes between $0.01\%$ and $10\%$ of the total population size. We simulate the degrees as $d_i \sim \text{Unif}(10, 1000)$. Finally, we let $c_k$ range evenly from the minimum possible value to the maximum possible value, such that the binomial probability is between 0 and 1 for all respondents. $\hat{N}_k / N$ ranges from $0.78$ to $1.46$. We implement Algorithm \ref{alg:adj} in a leave-one-out procedure, where we estimate the adjusted subpopulation sizes for each subpopulation sequentially, treating each successive subpopulation as unknown.

\begin{figure}[!t]
    \centering
    \begin{subfigure}{0.8\textwidth}
        \includegraphics[width=\textwidth]{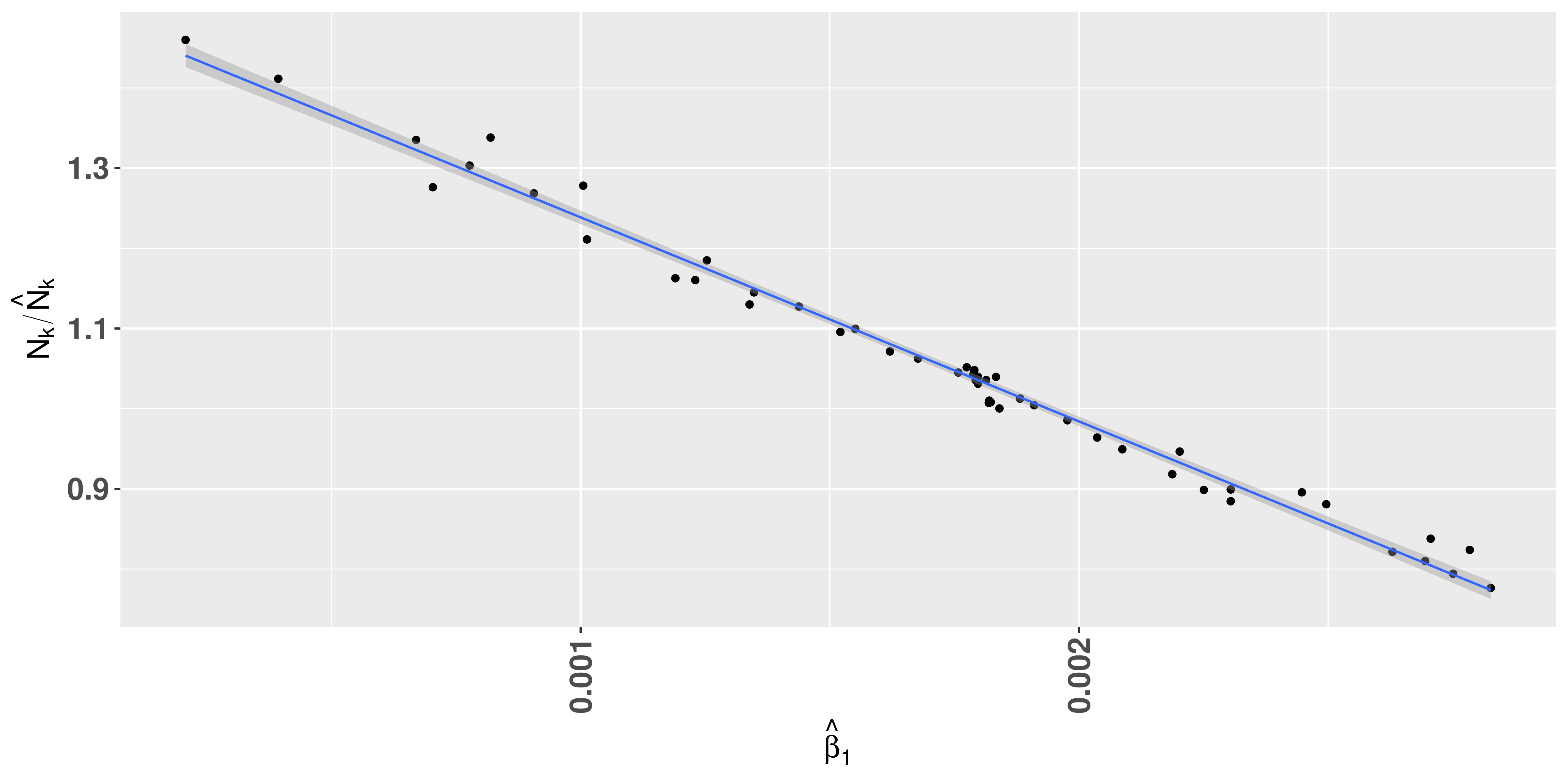}
        \caption{}
        \label{fig:dif_p_slopes}
    \end{subfigure}
        \begin{subfigure}{0.8\textwidth}
        \includegraphics[width=\textwidth]{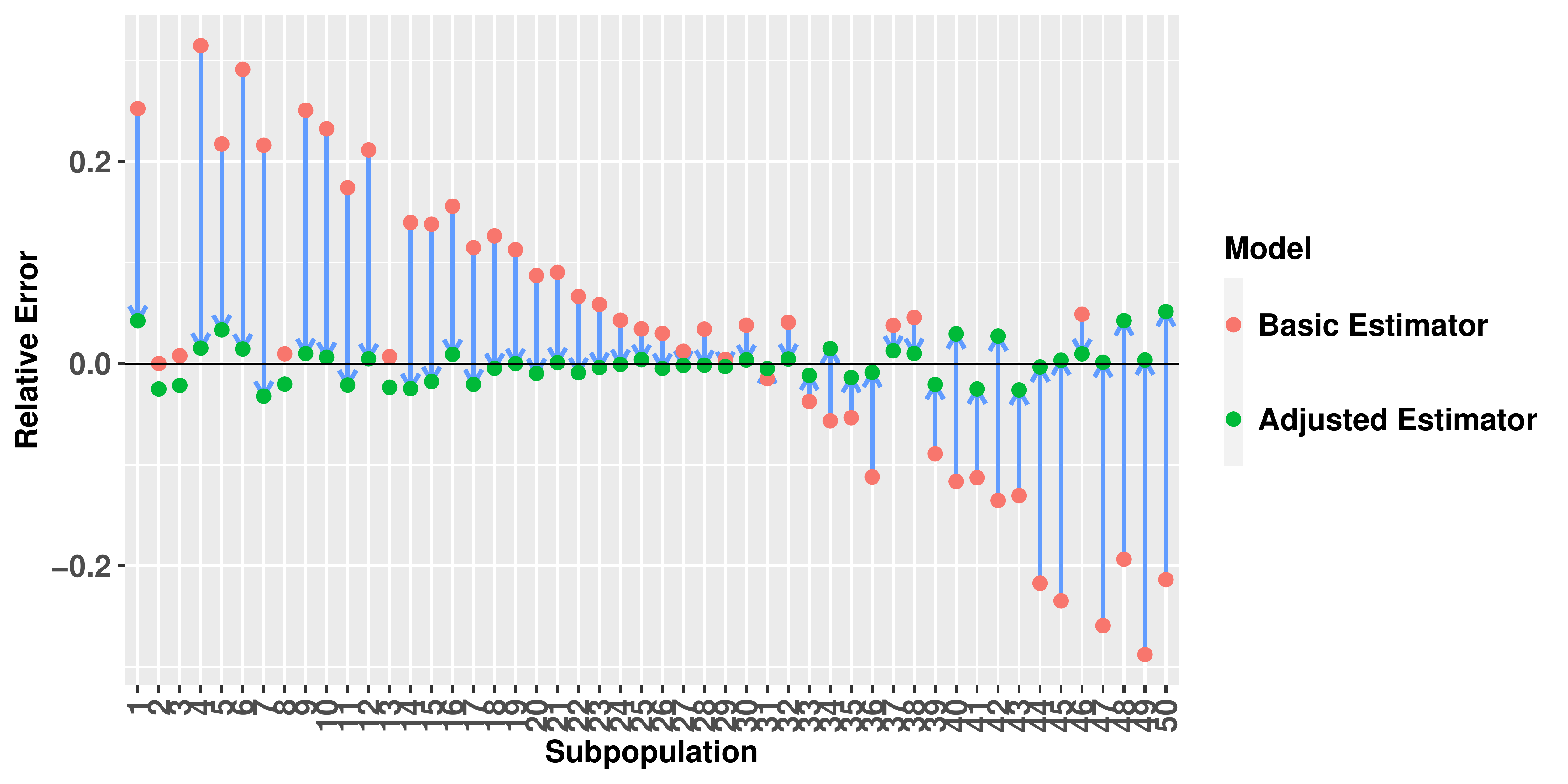}
        \caption{}
        \label{fig:dif_p_results}
    \end{subfigure}
    \caption{Empirical error of $N_k / \hat{N}_k$ plotted against the estimated first-stage slopes data simulated from a binomial model with different values of $p$ (a) and corresponding adjustment results.}
    \label{fig:dif_p}
\end{figure}

We plot the estimated slopes against $N_k / \hat{N}_k$ for the 50 subpopulations in Figure \ref{fig:dif_p_slopes}. The figure shows that the estimated slopes are still correlated with $N_k / \hat{N}_k$, although now with some additional noise.

The adjustment procedure results in a 88\% reduction in mean absolute percent error. Figure \ref{fig:dif_p_results} shows the performance of the two estimators across the 50 subpopulations. The adjusted estimator overwhelmingly outperforms the original estimator. To summarize, despite $p$ varying across different subpopulations, the proposed approach is still able to account for most of the estimator bias.

\end{document}